\documentclass[10pt,twocolumn,aps,prl,showpacs,superscriptaddress,floatfix]{revtex4-1}

\usepackage{graphicx}
\usepackage{amssymb}
\usepackage{amsmath}
\usepackage{hyperref}
\usepackage{color}

\newcommand{\mean}[1]{\langle #1 \rangle}

\newcommand{\ICFOAddress}{ICFO -- Institut de Ciencies Fotoniques, The Barcelona Institute of Science and Technology, 08860 Castelldefels, Barcelona, Spain}
\newcommand{\BilbaoAddress}{Department of Theoretical Physics, University of the Basque Country UPV/EHU, P.O. Box 644, E-48080 Bilbao, Spain}
\newcommand{\ICREAAddress}{ICREA -- Instituci\'{o} Catalana de Re{c}erca i Estudis Avan\c{c}ats, 08015 Barcelona, Spain}
\newcommand{\SiegenAddress}{Naturwissenschaftlich-Technische Fakult\"at, Universit\"at Siegen, Walter-Flex-Str. 3, D-57068 Siegen, Germany}
\newcommand{\HungaryAddress}{Wigner Research Centre for Physics, Hungarian Academy of Sciences, P.O. Box 49, H-1525 Budapest, Hungary}
\newcommand{\IkerbasqueAddress}{IKERBASQUE, Basque Foundation for Science, E-48013 Bilbao, Spain}
\begin{document}
\title{Quantum non-demolition measurement enables macroscopic Leggett-Garg tests}

\author{C. Budroni}
\thanks{These authors equally contributed to this work}
\affiliation{\SiegenAddress}

\author{G. Vitagliano}
\thanks{These authors equally contributed to this work}
\affiliation{\BilbaoAddress}

\author{G. Colangelo }
\thanks{These authors equally contributed to this work}
\affiliation{\ICFOAddress}
\email[]{giorgio.colangelo@icfo.es}

\author{R.J.~Sewell}
\affiliation{\ICFOAddress}

\author{O. G\"uhne}
\affiliation{\SiegenAddress}

\author{G. T\'{o}th}
\affiliation{\BilbaoAddress}
\affiliation{\IkerbasqueAddress}
\affiliation{\HungaryAddress}

\author{M.W.~Mitchell}
\affiliation{\ICFOAddress}
\affiliation{\ICREAAddress}

\date{ \today}

\begin{abstract}
We show how a test of macroscopic realism based on Leggett-Garg inequalities (LGIs) can be performed in a macroscopic system. 
Using a continuous-variable approach, we consider quantum non-demolition (QND) measurements applied to atomic ensembles undergoing magnetically-driven coherent oscillation. 
We identify measurement schemes requiring only Gaussian states as inputs and giving a significant LGI violation with realistic experimental parameters and imperfections. 
The predicted violation is shown to be due to true quantum effects rather than to a classical invasivity of the measurement. 
Using QND measurements to tighten the ``clumsiness loophole'' forces the stubborn macrorealist to re-create quantum back action in his or her account of measurement. 
\end{abstract}
\maketitle

Making analogies to Bell inequalities~\cite{BellP1964}, Leggett and Garg (LG)~\cite{LeggettGargPRL1985} proposed a test for quantum behavior of macroscopic systems undergoing coherent evolution. 
The resulting Leggett-Garg inequalities (LGIs) aim to distinguish a hypothesized philosophical position of {\em macrorealism} (MR) from quantum physics, and ultimately to test this position against nature. 
The MR position holds that arbitrarily low-disturbance measurements should be possible, contradicting the Heisenberg uncertainty principle. 

To make manifest the LG ideas, several experiments have tested LGIs. 
Nearly all experiments have used microscopic systems, including single photons~\cite{GogginPNAS2011,xuSR2011,DresselPRL2011,SuzukiNJP2012}, a single photon stored in a macroscopic quantum memory~\cite{ZhouPRL2015}, defects in diamonds~\cite{WaldherrPRL2011,GeorgePNAS2013}, nuclear spins~\cite{AthalyePRL2011,SouzaNJP2011,KneSimGau12,KatiyarPRA2013}, and cold atoms~\cite{RobensPRX2015}.
See Ref.~\cite{EmariNoriRPR2014} for a review. 
To date, two experiments have tested LGIs on macroscopic systems outside the single-excitation regime: those by Palacios-Laloy {\em et al.}~\cite{PalaciosNatPhot2010} and
Groen {\em et al.}~\cite{GroenPRL2013}. These experiments used superconducting qubits and showed a significant violation of a LG-like inequality for weak measurements \cite{RuskovPRL2006,WilliamsPRL2008}.  
 
Because continuous weak measurements record a system oscillating between two conjugate variables, they perturb both variables during the multi-cycle measurement. This guarantees a disturbance and opens a ``clumsiness loophole''~\cite{WildeFP2012}: a macro-realist can interpret the LGI violation as caused by imperfect (from the MR perspective) measurements. 
As argued by Wilde and Mizel (WM)~\cite{WildeFP2012}, the clumsiness loophole cannot be closed, but one can force the macrorealist to retreat to unlikely scenarios in which the clumsiness is imperceptible except in the LGI test. 
WM considered ideal projective measurements, which can be well approximated only for microscopic physical quantities. Quantum non-demolition (QND) measurement is a practical alternative suitable for macroscopic quantities.
A QND measurement has both measurement uncertainty and disturbance (of the measured variable) near or below the standard quantum limit~\cite{GrangierN1998}. 
Originally proposed to detect mechanical oscillations in gravitational wave detectors~\cite{BraginskyS1980}, a strictly-defined QND measurement has been demonstrated in optical~\cite{GrangierN1998} and in atomic~\cite{SewellNatPhot2013} systems.

Here we show that, in contrast to previous approaches, QND measurements 
can test a macroscopic system against a true LGI, i.e. absent additional assumptions. The approach closely resembles the original LG proposal and maximally tightens the clumsiness loophole. 
Using the collective quantum variable formalism~\cite{MadsenPRA2004,ColangeloNJP2013}, we predict a violation for realistic experimental parameters~\cite{KubasikPRA2009,ColangeloNJP2013},  with the possibility of a straightforward extension of the analysis to different macroscopic systems, up to everyday-life scales \cite{AbbottNJP2009}.
This disproves a well-known conjecture \cite{KoflerPRL2008} that LGI violation in a 
macroscopic system requires high computational complexity and, thus, that suitable systems are 
unlikely to exist in nature (see also Refs. \cite{WangPRA2013, JeongPRL2014}). It 
also extends previous studies of QND-measurement-based LG tests \cite{CalarcoPRA1995,CalarcoAPB1997} 
that found no LGI violation with three-measurement protocols, and it explicitly shows a fundamental 
difference between temporal (LG) and spatial (Bell) non-classicality in the macroscopic limit 
\cite{Ramanathan2007}.
Our calculation method allows a clear discrimination between incidental disturbances from, 
e.g., spontaneous scattering, and essential disturbance due to quantum back-action. The 
clumsiness loophole can be tightened as per WM, and doing so forces the macrorealist to a position strongly resembling quantum mechanics.

\begin{figure}
\includegraphics[width=\columnwidth,clip]{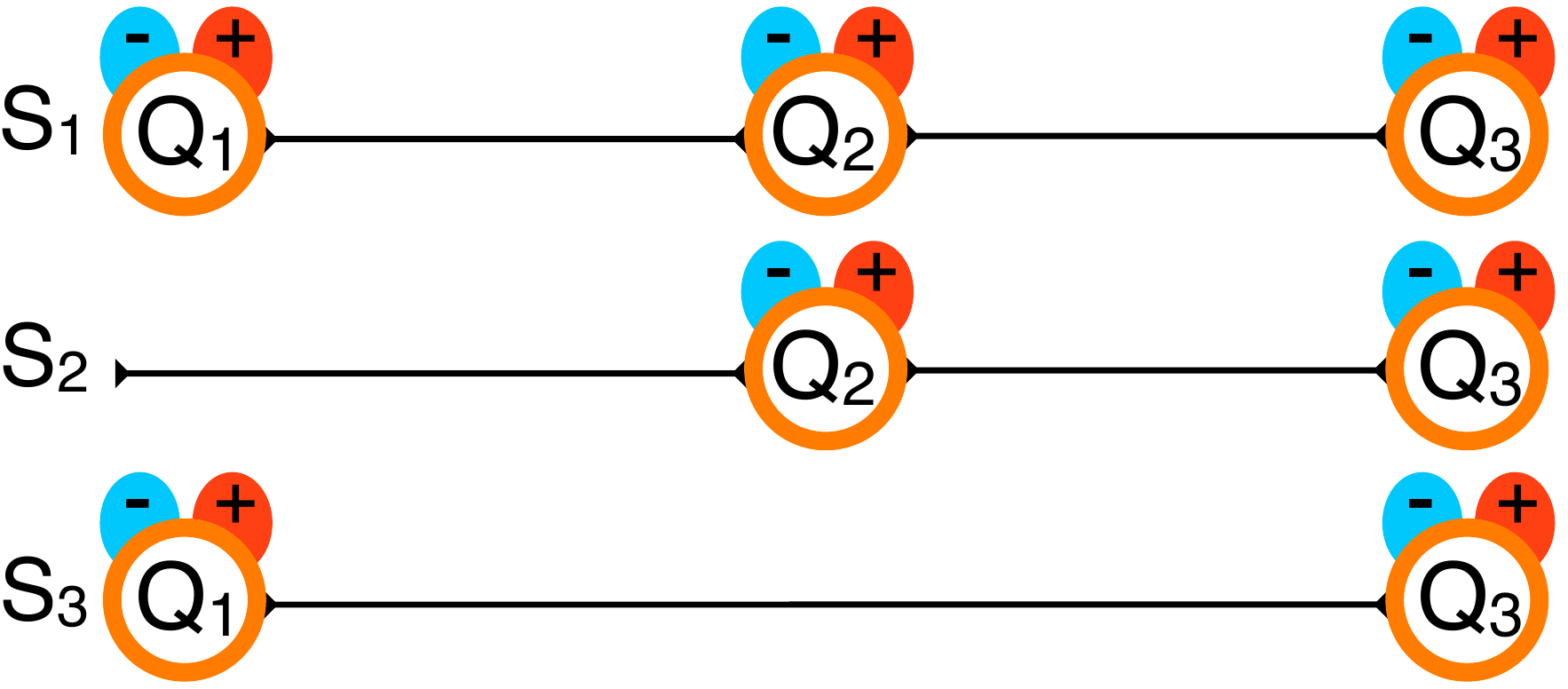}
\caption{
Schematic representation of the Leggett-Garg test. 
{An observable $Q$ is measured at different times $t_i$ (represented along the horizontal axis), giving results $Q_i$. 
Macrorealism assumes non-invasive measurements, with the consequence that correlations, e.g., $C_{13} = \mean{Q_1 Q_3}$, are equal, independently of which sequence was performed to obtain them. $S_1$ and $S_3$, which differ by the presence or absence of $Q_2$, give the same $C_{13}$ in macrorealism but not in quantum mechanics. LGIs can detect macrorealism violations using experimentally observed correlations. }
 }\label{lg3}
\end{figure} 
{\it Leggett-Garg inequalities---}The simplest LGI~\cite{LeggettGargPRL1985, EmariNoriRPR2014} considers two-time correlation functions $C_{ij}=\mean{Q(t_i)Q(t_j)}$, for an observable $Q$ measured at times $t_i$, and giving discrete values $Q(t) = \pm 1$~\cite{LeggettGargPRL1985} or else limited to the range $Q(t) \in [-1,1]$~\cite{RuskovPRL2006} (cf. Fig.~\ref{lg3}). 
Under MR, the correlations obey
\begin{equation}\label{LGdef}
K_3\equiv C_{21}+C_{32}+C_{31}+1\geq 0,
\end{equation}
while a quantum system can show $K_3=-1/2$ in the case of a two-level system~\cite{LeggettGargPRL1985}, and can reach the algebraic bound $K_3=-2$ for an infinite number of levels~\cite{BudroniPRL2013,BudroniPRL2014}.
Equation (\ref{LGdef}) can be generalized to a class of inequalities involving $n$ measurements in time~\cite{AvisPRA2010}, namely,
\begin{equation}\label{K_n}
K_n \equiv \sum_{1\leq j<i\leq n} \mean{Q_i Q_j} + \left\lfloor \frac{n}{2}\right \rfloor\geq 0,
\end{equation}
where $\left\lfloor n \right \rfloor$, denotes the integer part of $n$.
Such inequalities correspond to facets of the Leggett-Garg polytope, and therefore provide optimal discrimination of non-classical correlations~\cite{Pitowsky89b,DL97}. 

{\it QND measurement---} An ideal QND measurement \cite{BraginskyS1980,GrangierN1998} is an indirect measurement of a variable $P_S$ performed by coupling the ``system'' (S) to a ``meter'' (M) via an interaction Hamiltonian $H_\mathrm{int} = P_{\rm S} P_{\rm M}$, where $P_{\rm S,M}$ are the conjugate momenta to $X_{\rm S,M}$, respectively.  The coupling imprints information about $P_{\rm S}$ on $X_{\rm M}$, without disturbing $P_{\rm S}$, which commutes with $H_I$. 
A strong, direct measurement is then made on $X_{\rm M}$, providing information about $P_{\rm S}$.  This leaves the system in a state with reduced uncertainty in $P_{\rm S}$. In contrast the QND interaction produces a back action on the conjugate variable $X_{\rm S}$, increasing its uncertainty.

Detailed and accurate models of QND measurements, including realistic models for measurement-induced disturbance, have been developed for atomic ensembles probed by near-resonant light~\cite{ColangeloNJP2013, BaragiolaPRA2014}. 
We use the collective quantum variables formalism introduced in Ref.~\cite{MadsenPRA2004} and expanded in Refs.~\cite{KoschorreckJPB2009,ColangeloNJP2013}. The same formalism can be applied to other macroscopic systems \cite{MadsenPRA2004}.

The atomic system we consider consists of $N_A$ spin-1 atoms, described by the collective spin vector $\mathbf{J}$ with components $J_k\equiv \sum_{l} j_k^{(l)}$, where ${\mathbf{j}^{(l)}}$ is the total angular momentum of the $l$th atom. 
The probe light, or ``meter'' consists of pulses of $N_L$ photons described by the Stokes vector $\mathbf{S}^{(i)}$ for the $i$th pulse with components $S_k = \frac{1}{2}(a_L^\dagger, a_R^\dagger) \sigma_k (a_L, a_R)^T$ where $\sigma$ are the Pauli matrices. 
The system plus meter are described by the vector of observables $\mathbf{V} = (\mathbf{J},\mathbf{S}^{(1)}, \ldots, \mathbf{S}^{(n)}) $, where $n$ is the total number of light pulses. 

The initial state is fully $x$-polarized, i.e. with $\mean{{J_x}} = N_A$ and $\mean{{S^{(i)}_x}} = N_L/2$. The angular momentum components $J_y, J_z$ and the Stokes components $S_y, S_z$ have zero mean and, due to the large numbers, $N_A \sim 10^6$ and $N_L \sim 10^8$, are Gaussian distributed to a good approximation. In this sense, we use a macroscopic number of photons to perform a measurement on  a macroscopic number of atoms.
We can thus describe the full state using the average $\mean{\mathbf{V}}$ and the covariance matrix~\cite{GiedkePRA2002,MadsenPRA2004,KoschorreckJPB2009,ColangeloNJP2013} 
\begin{equation}\label{covvardef}
\Gamma_{ij} \equiv \frac 1 2 \mean{V_i V_j +V_j V_i} - \mean{V_i}\mean{V_j} .
\end{equation}
Free dynamics under a magnetic field along the $x$-direction produces the evolution 
\begin{equation}
\begin{aligned}
\mean{\mathbf{J}}\mapsto \mean{\mathbf{J}_{\theta}} &= R_{x}(\theta)\mean{\mathbf{J}} , \\
\Gamma_{\mathbf{J}} \mapsto \Gamma_{\mathbf{J}_{\theta}} &= R_{x}(\theta) \Gamma_{\mathbf{J}} R_{x}(\theta)^T ,
\end{aligned}
\end{equation}
where $\theta\equiv\kappa B \Delta t$  is the rotation angle of the atoms in the time $\Delta t$ given the coupling constant $\kappa\equiv -\mu_{\rm B} g_{\rm F}$ between the atoms and a magnetic field with amplitude $B$, if $\mu_{\rm B}$ is the Bohr magneton and $g_{\rm F}$ the Land\'e factor; $\Gamma_{\mathbf{J}}$ refers to just the atomic part of the covariance matrix, and $R_{x}(\theta)$ is the matrix describing rotation about the $J_x$ axis.
A measurement consists of passing a pulse of light, which is short relative to the Larmor precession time, through the atoms. Faraday rotation, produced by a ``QND'' interaction Hamiltonian $ H_{\mathrm{int}}=g S_z J_z $,imprints the instantaneous value of $J_z$ on the light, described by the relations (in the small angle approximation)
\begin{subequations}
\label{eq:main}
\begin{eqnarray}
S_y^{\rm (out)} &= S_y^{\rm (in)} + g J_z^{\rm (in)}S_x^{\rm (in)} ,\label{outprelV}\\
J_y^{\rm (out)} &= J_y^{\rm (in)} + g J_x^{\rm (in)}S_z^{\rm (in)} , \label{outprelGamma}
\end{eqnarray}
\end{subequations}
where $g$ is a coupling constant. $J_z$ and $S_z$, which commute with $H_I$, are unchanged. 

The required Hamiltonian has been achieved by dynamical decoupling~\cite{KoschorreckPRL2010b,SewellNatPhot2013} and by two-color probing~\cite{SaffmanPRA2009}. 

The linear operator relations of Eqs.~(\ref{outprelV}) and~(\ref{outprelGamma}) transform the mean and covariance matrix as 
\begin{equation}
\begin{aligned}
\mean{\mathbf{V}}&\mapsto M_Q\mean{\mathbf{V}} , \\
\Gamma&\mapsto M_Q\Gamma M_Q^T,
\end{aligned}
\end{equation}
where the matrix $M_Q$ has the form~\cite{MitchellNJP2012} 
\begin{equation}
\label{eq:M_matrix}
M_Q=\left(
\begin{matrix}
\mathcal{A}&\mathcal{B}_{\rm at}\\
\mathcal{B}_{\rm l}&\mathcal{L}
\end{matrix}
\right).
\end{equation}
The matrix elements $\mathcal{B}_{\rm at}$ and $\mathcal{B}_{\rm l}$ represent the back action of the interaction on the atoms ($\mathbf{J}$) and light ($\mathbf{S}$), respectively, and are practically given by the terms multiplying $g$ in Eqs.~(\ref{outprelV}), (\ref{outprelGamma}).
In a realistic situation, i.e. for finite optical depth (OD), we must take into account loss and decoherence due to off-resonant scattering of the QND probe light.
As discussed in Ref.~\cite{ColangeloNJP2013}, if a fraction $1-\chi$ of the $N_A$ atoms scatters a photon, this alters the variances of the quantum components $y$ and $z$ as
\begin{equation}\label{loss}
\Gamma_{\mathbf{J}}\mapsto \chi^2 \Gamma_{\mathbf{J}}+N_A(1-\chi)(\frac{\chi}{2}+\frac{2}{3})\openone . 
\end{equation}
Note that for constant $g$, $\chi \rightarrow 1$ (i.e., no scattering noise) as OD $\rightarrow \infty$. Repeated application of these transformation rules gives $\mean{\mathbf{J}}$ and $\Gamma_{\mathbf{J}}$ describing the now-correlated system after all pulses have traversed the ensemble. Through Eq. ~(\ref{outprelV}), the meter variables $S_y^{(i)}$ can be taken to represent the correlated measurement outcomes on the system observable $J_z(t_i)$ at times $t_i$.

\begin{figure}
\includegraphics[width=\columnwidth,clip]{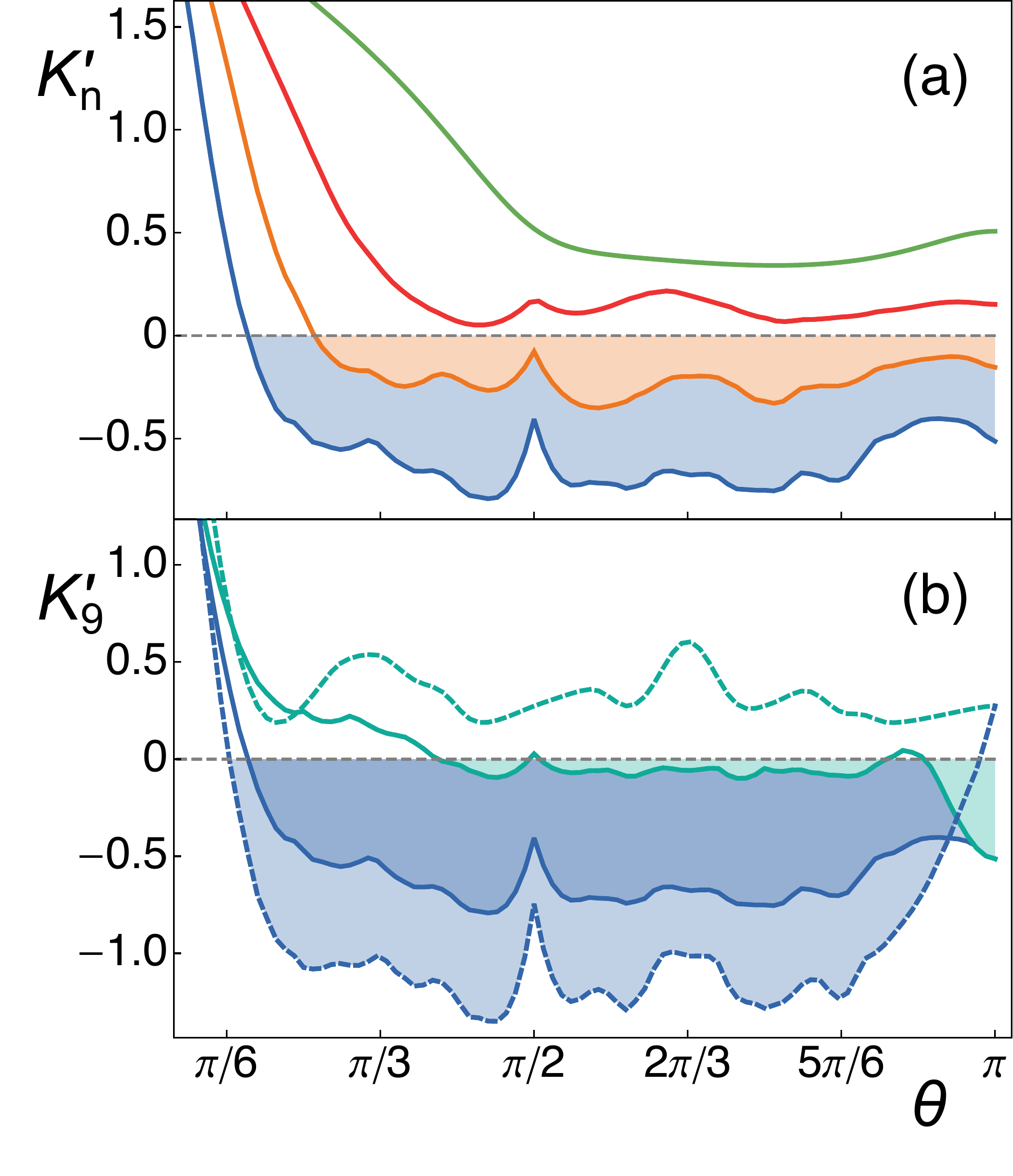}
\caption{ 
Numerical evaluation of the reduced LGIs as a function of $\theta$. 
(a) $K'_n$ for (from top to bottom) $n=3,5,7,9$ in the presence of scattering.
(b) Numerical evaluation of $K'_{9}$. The two lower blue (upper green) curves are results with (without, i.e. $\mathcal{B}_{\rm at}=0$) back action in Eq.~(\ref{eq:M_matrix}).
Solid (dashed) lines show results with (without) scattering. 
All plots are obtained using the same parameters, taken from Ref.~\cite{ColangeloNJP2013} (see text).
}\label{plotKndiscsc}
\end{figure}

{\it Application to LG inequalities---}The $n$ outcomes ${\bf y}\equiv(y_1, \ldots, y_n) =(S_y^{(1)},\ldots,S_y^{(n)})$ are distributed according to the Gaussian probability density function 
\begin{equation}
\Pr({\bf y})=G^{(n)}_{\Gamma_\mathcal{Y}}({\bf y})=\frac{\exp{\left[- ({\bf y} - \boldsymbol{\mu})^T \Gamma^{-1}_\mathcal{Y} ({\bf y} - \boldsymbol{\mu}) \right]}}{\sqrt{(2\pi)^n \det \Gamma_\mathcal{Y}}} ,
\end{equation}
with mean $\boldsymbol{\mu}=0$, where $\Gamma_\mathcal{Y}$ is the covariance matrix describing $\mathbf{y}$~\footnote{It is worth recalling that, in this approximation, the outcomes have a continuous and unbounded spectrum $y \in \mathbb R$, although the total angular momentum is discrete and bounded.}. 

\newcommand{\sign}{{\rm sgn}}

As in the original LG article, we generate a dichotomic variable $Q(t_i) \equiv \sign(y_i)$ \footnote{The literature contains different notions of coarse-graining \cite{KoflerPRL2008, 
 WangPRA2013, JeongPRL2014,BudroniPRL2014}. We note that our procedure of splitting the range of $y$ into two regions is precisely that of the original LG proposal \cite{LeggettGargPRL1985}. 
}. The correlators $C_{ij}$ can be evaluated from the $2\times 2$ covariance matrix $\Gamma_{\mathcal{Y}_{ij}}=
\left(
\begin{smallmatrix}
 A&B\\
 B&C 
\end{smallmatrix}
\right)$ 
obtained as the submatrix of $\Gamma_\mathcal{Y}$ describing the measured pulses $S_y^{(i)},S_y^{(j)}$.
In particular $C_{ij}=\left(1-2\alpha/ \pi \right) \sign(B) $, where $\alpha= \arctan (\sqrt{AC/B^2-1} )$.

{\it Results---}Predicted LGI outcomes are shown in Fig. ~\ref{plotKndiscsc}, where we evaluate sequences with $n=3,5,7$, and $9$ measurements.
To directly compare these cases, we consider a reduced LG parameter $K'_n\equiv K_n/\left\lfloor \frac{n}{2}\right \rfloor$, noting that $K'_3=K_3$.
We evaluate $K'_{3-9}$, taking into account decoherence and losses as in Eq.(\ref{loss}).
For simplicity, we consider the case of equally delayed measurements, i.e. {a rotation angle $\theta$ between each of the $n$ measurements}. 
Realistic parameters are used: $g=10^{-7}$, $N_A=10^6$, $N_L=5\times 10^8$, and $\chi=\exp[-\eta N_L]$, where $\eta=0.5 \times 10^{-9}$~\cite{ColangeloNJP2013}. 
No violation is seen with an $n=3$ protocol; a violation is seen with $n=5$, but only with very low $\eta$, below current experimental values (not shown). 
For $n=7$ and $n=9$, the LGI violation is achievable with realistic parameters [see Fig.~\ref{plotKndiscsc}]. 
In Fig.~\ref{plotKndiscsc}(b) we compare the $n=9$ case with and without {loss and noise introduced due to off-resonant scattering}. 
Note that for most $\theta$, the effect of unwanted scattering is to reduce the observed violation. 
In contrast, for $\theta\approx\pi$, scattering increases the violation, or can create an apparent violation that is absent for an ideal measurement (i.e. with $\eta = 0$).

The above tests involve a large number of correlation terms (e.g., $21$ for computing $K_7$).
We can considerably simplify the protocol and reduce the number of measurement sequences by considering just a triple $\{Q_a, Q_b, Q_c\}$, extracted out of the $n$-measurement scheme, and the corresponding correlators, namely
\begin{equation}\label{k3abc}
K_3 = C_{ab} + C_{bc} + C_{ac} + 1 \geq 0 .
\end{equation}

\begin{figure}
\includegraphics[width=\columnwidth]{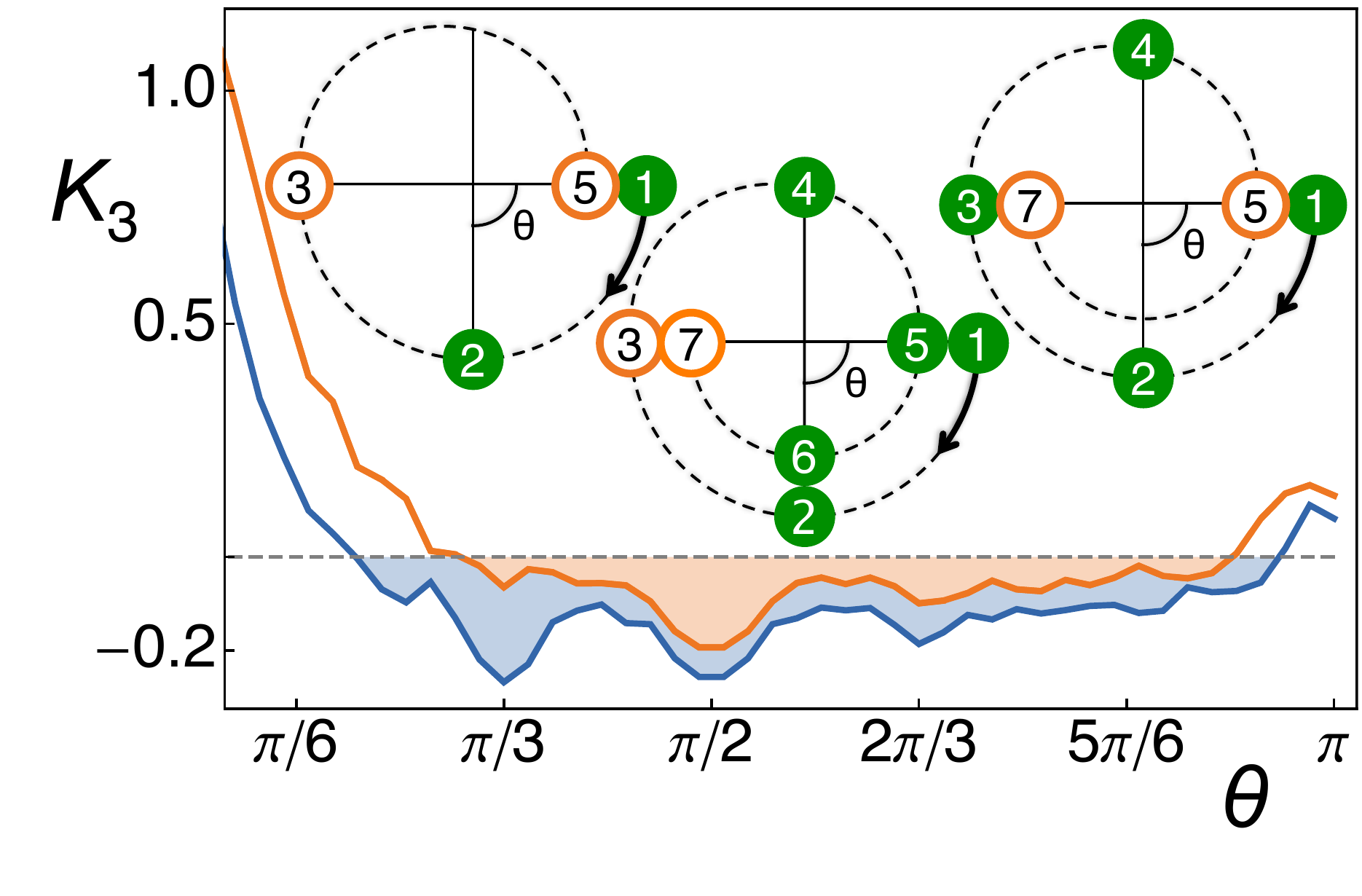}
\caption{Three-point LGI violations within longer measurement sequences. 
Upper and lower curves show $K_3$ versus $\theta$ for optimized seven- and nine-measurement sequences, respectively. Both plots are obtained for $N_L= 5\times 10^8$ including scattering and loss effects. Spirals illustrate optimal sequences in the seven-measurement protocol with delay $\theta=\frac{\pi}{2}$, for which $K_3 = C_{35} + C_{37} + C_{57} + 1$. Hollow orange circles indicate measurements used to compute correlators in $K_3$, filled green circles indicate measurements performed but discarded. }
\label{figK3from7}
\end{figure}

We compute the best achievable $K_3$ optimizing over all possible triples and all possible sequences, since from a macrorealist perspective possible additional measurements have no effect. 
The results are plotted in Fig.~\ref{figK3from7}, where it can be seen that a violation of Eq.~(\ref{k3abc}) is obtainable, especially around the points $\theta=\tfrac{\pi} 2$ and $\theta=\tfrac{\pi} 3$.
The optimal sequences of seven measurements for $\theta=\tfrac{\pi} 2$ are also depicted in Fig.~\ref{figK3from7}. 
For an ideal QND measurement we should get $C_{35}=C_{57}=-1$, with $C_{37}<1$ due to the various discarded measurements made between $Q_3$ and $Q_7$ that decorrelate the two measurements and give rise to the LGI violation. 
Compared to the seven-point measurement of Fig.~\ref{plotKndiscsc}, this protocol shows less violation but requires fewer measurement sequences {and involves calculation of a simpler correlation function, potentially making it more robust in the presence of experimental uncertainties.}

{\it Classical versus quantum effects---} Within a quantum interpretation, we can ask whether the violation of LGIs witnesses a genuine quantum effect or whether it is due to the classical invasivity (clumsiness) of the measurement.
There are two ingredients that contribute to the violation: the  {\it scattering} and the {\it quantum back action} of the measurement on $J_y$.
The violation around $\theta=\pi$ can be easily explained in terms of the classical invasivity of the measurement: measurements at angles $k\pi$, which should be perfectly correlated or anticorrelated, are decorrelated due to scattering effects.
On the other hand, the {quantum back action} is a genuinely quantum effect, required by the Heisenberg uncertainty principle.

Our formalism allows us to distinguish between these two contributions, and we do so by simulating a QND measurement where the effect of the quantum back action is ``turned off'', i.e. $\mathcal{B}_{\rm at}=0$ corresponding to $J_y^{\rm (out)}=J_y^{\rm (in)}$ in the input-output relations of Eqs. ~(\ref{outprelV}) and ~(\ref{outprelGamma}).
 
The results in this last case are shown in the green upper curves of the bottom of 
Fig.~\ref{plotKndiscsc}(b). These results show that the violation genuinely comes from the 
quantum back action effect in most of the cases. Scattering becomes important only at some specific 
phases, and is responsible for a significant violation only for $\theta$ approaching $\pi$.

{\it Tightening the clumsiness loophole---}
WM suggest performing, in addition to the LGI test, auxiliary measurement sequences that prove the individual measurements are non-disturbing~\cite{WildeFP2012}, and describe appropriate sequences for projective measurements on qubits. 
Projective measurements are unrealistic in the macroscopic context, however. We now show how even non-projective QND measurements can be proven to be non-disturbing; see also~\cite{MitchellNJP2012,SewellNatPhot2013}. 

\newcommand{\var}{{\rm var}}

Consider two identical, non-destructive measurements in rapid succession, i.e, with no system evolution in between. 
If the statistics of the first and second measurements agree, the first cannot have disturbed the system. 
We illustrate this with linear measurements of $J_z$ with known gain $g$, described as $S_y^{(\rm out)} = n + g J_z^{(\rm in)} $ and $J_z^{(\rm out)} = J_z^{(\rm in)} + d$, where the random variables $n$ and $d$ are the readout noise and the disturbance to $J_z$, respectively.
Considering two identical measurements in quick succession, it is easily shown that $\langle S_y^{(2)} -S_y^{(1)} \rangle = \langle d \rangle$ and $\var(S_y^{(2)}) -\var(S_y^{(1)}) = g^2 \var(d)$, which provide ready quantifications of $\langle d \rangle$ and $\var(d)$. 
Both of these approach zero for QND measurements in the limit of high OD (the ``ideal case'' of Fig. ~\ref{plotKndiscsc}) 
\footnote{The vanishing of $d$ implies low damage, not high measurement resolution. The case of finite optical depth will be treated in a separate publication.}.
It is thus possible to demonstrate to a macrorealist that the QND measurement does not disturb $J_z$. 
 
Combined with the LGI violation, this puts the macrorealist in a tight spot, requiring some kind of ``colluding measurements'' (in the words of WM) to explain the LGI. 
To remain within the framework of realistic explanation, the macrorealist must believe there is a condition of the system after the QND measurement. 
This clearly involves a change in the state but not of $J_z$; some other, orthogonal variable must change. 
Moreover, this disturbance must give rise to a LGI violation, so it must be a variable that, in time, rotates into $J_z$. 
Given that the magnetic rotation is about the $x$ axis, $J_x$ does not rotate into $J_z$, and the only option is that the measurement disturbs $J_y$. 
Being orthogonal to $J_z$, this disturbance does not show up in the quickly repeated measurements of the auxiliary sequences, but it becomes visible later as the state evolves, leading to the LGI violation. 
Remarkably, this macrorealist explanation reproduces precisely, if qualitatively, the quantum mechanical explanation. 
The macrorealist must re-invent quantum back action to describe the alleged ``clumsiness.''

{\it Conclusions and discussion---} 
We have shown that quantum non-demolition measurements allow true Leggett-Garg inequalities to be tested in macroscopic systems. 
Protocols involving simple state preparation and as few as five measurements can violate a generalized LGI, and the degree of violation grows with the number of measurements. 
Using covariance matrix simulations, we can distinguish violations due to quantum back action from violations due to incidental but unavoidable effects such as incoherent scattering.
We show how QND measurements can be used to tighten the ``clumsiness loophole'' in a macroscopic system and force the macrorealist to a position closely resembling quantum mechanics. The LG test strategy described here uses only Gaussian states and Gaussian measurements, and can foreseeably be applied to very large objects such as gravitational wave interferometer mirrors \cite{AbbottNJP2009}.

{\it Acknowledgements---}
The authors thank A. Asadian, R. Di Candia, C. Emary, M. Kleinmann, F. Mart\' in Ciurana, S. Palacios, F. E. S. Steinhoff, and T. Tufarelli for stimulating discussions.
The work was supported by ERC starting grants GEDENTQOPT and AQUMET, CHIST-ERA Project QUASAR, MINECO Projects No. FIS2012-36673-C03-03 and No. FIS2011-23520, National Research Fund of Hungary OTKA (Contract No. K83858), Basque Government Project No. IT4720-10, Marie Curie CIG 293993/ENFOQI, the FQXi Fund (Silicon Valley Community Foundation), the DFG, and Fundaci\'{o} Privada Cellex. 


~
\bibliographystyle{apsrev4-1no-url}
\bibliography{LG}

\begin{thebibliography}{47}%
\makeatletter
\providecommand \@ifxundefined [1]{%
 \@ifx{#1\undefined}
}%
\providecommand \@ifnum [1]{%
 \ifnum #1\expandafter \@firstoftwo
 \else \expandafter \@secondoftwo
 \fi
}%
\providecommand \@ifx [1]{%
 \ifx #1\expandafter \@firstoftwo
 \else \expandafter \@secondoftwo
 \fi
}%
\providecommand \natexlab [1]{#1}%
\providecommand \enquote  [1]{``#1''}%
\providecommand \bibnamefont  [1]{#1}%
\providecommand \bibfnamefont [1]{#1}%
\providecommand \citenamefont [1]{#1}%
\providecommand \href@noop [0]{\@secondoftwo}%
\providecommand \href [0]{\begingroup \@sanitize@url \@href}%
\providecommand \@href[1]{\@@startlink{#1}\@@href}%
\providecommand \@@href[1]{\endgroup#1\@@endlink}%
\providecommand \@sanitize@url [0]{\catcode `\\12\catcode `\$12\catcode
  `\&12\catcode `\#12\catcode `\^12\catcode `\_12\catcode `\%12\relax}%
\providecommand \@@startlink[1]{}%
\providecommand \@@endlink[0]{}%
\providecommand \url  [0]{\begingroup\@sanitize@url \@url }%
\providecommand \@url [1]{\endgroup\@href {#1}{\urlprefix }}%
\providecommand \urlprefix  [0]{URL }%
\providecommand \Eprint [0]{\href }%
\providecommand \doibase [0]{http://dx.doi.org/}%
\providecommand \selectlanguage [0]{\@gobble}%
\providecommand \bibinfo  [0]{\@secondoftwo}%
\providecommand \bibfield  [0]{\@secondoftwo}%
\providecommand \translation [1]{[#1]}%
\providecommand \BibitemOpen [0]{}%
\providecommand \bibitemStop [0]{}%
\providecommand \bibitemNoStop [0]{.\EOS\space}%
\providecommand \EOS [0]{\spacefactor3000\relax}%
\providecommand \BibitemShut  [1]{\csname bibitem#1\endcsname}%
\let\auto@bib@innerbib\@empty
\bibitem [{\citenamefont {Bell}\ \emph {et~al.}(1964)\citenamefont {Bell} \emph
  {et~al.}}]{BellP1964}%
  \BibitemOpen
  \bibfield  {author} {\bibinfo {author} {\bibfnamefont {J.~S.}\ \bibnamefont
  {Bell}} \emph {et~al.},\ }\href@noop {} {\bibfield  {journal} {\bibinfo
  {journal} {Physics}\ }\textbf {\bibinfo {volume} {1}},\ \bibinfo {pages}
  {195} (\bibinfo {year} {1964})}\BibitemShut {NoStop}%
\bibitem [{\citenamefont {Leggett}\ and\ \citenamefont
  {Garg}(1985)}]{LeggettGargPRL1985}%
  \BibitemOpen
  \bibfield  {author} {\bibinfo {author} {\bibfnamefont {A.~J.}\ \bibnamefont
  {Leggett}}\ and\ \bibinfo {author} {\bibfnamefont {A.}~\bibnamefont {Garg}},\
  }\href {\doibase10.1103/PhysRevLett.54.857} {\bibfield  {journal} {\bibinfo
  {journal} {Phys. Rev. Lett.}\ }\textbf {\bibinfo {volume} {54}},\ \bibinfo
  {pages} {857} (\bibinfo {year} {1985})}\BibitemShut {NoStop}%
\bibitem [{\citenamefont {Goggin}\ \emph {et~al.}(2011)\citenamefont {Goggin},
  \citenamefont {Almeida}, \citenamefont {Barbieri}, \citenamefont {Lanyon},
  \citenamefont {O'Brien}, \citenamefont {White},\ and\ \citenamefont
  {Pryde}}]{GogginPNAS2011}%
  \BibitemOpen
  \bibfield  {author} {\bibinfo {author} {\bibfnamefont {M.~E.}\ \bibnamefont
  {Goggin}}, \bibinfo {author} {\bibfnamefont {M.~P.}\ \bibnamefont {Almeida}},
  \bibinfo {author} {\bibfnamefont {M.}~\bibnamefont {Barbieri}}, \bibinfo
  {author} {\bibfnamefont {B.~P.}\ \bibnamefont {Lanyon}}, \bibinfo {author}
  {\bibfnamefont {J.~L.}\ \bibnamefont {O'Brien}}, \bibinfo {author}
  {\bibfnamefont {A.~G.}\ \bibnamefont {White}}, \ and\ \bibinfo {author}
  {\bibfnamefont {G.~J.}\ \bibnamefont {Pryde}},\ }\href
  {\doibase10.1073/pnas.1005774108} {\bibfield  {journal} {\bibinfo  {journal}
  {Proceedings of the National Academy of Sciences}\ }\textbf {\bibinfo
  {volume} {108}},\ \bibinfo {pages} {1256} (\bibinfo {year}
  {2011})}\BibitemShut {NoStop}%
\bibitem [{\citenamefont {Xu}\ \emph {et~al.}(2011)\citenamefont {Xu},
  \citenamefont {Li}, \citenamefont {Zou},\ and\ \citenamefont
  {Guo}}]{xuSR2011}%
  \BibitemOpen
  \bibfield  {author} {\bibinfo {author} {\bibfnamefont {J.-S.}\ \bibnamefont
  {Xu}}, \bibinfo {author} {\bibfnamefont {C.-F.}\ \bibnamefont {Li}}, \bibinfo
  {author} {\bibfnamefont {X.-B.}\ \bibnamefont {Zou}}, \ and\ \bibinfo
  {author} {\bibfnamefont {G.-C.}\ \bibnamefont {Guo}},\ }\href
  {\doibase10.1038/srep00101} {\bibfield  {journal} {\bibinfo  {journal}
  {Scientific Reports}\ }\textbf {\bibinfo {volume} {1}},\ \bibinfo {pages}
  {101 EP } (\bibinfo {year} {2011})}\BibitemShut {NoStop}%
\bibitem [{\citenamefont {Dressel}\ \emph {et~al.}(2011)\citenamefont
  {Dressel}, \citenamefont {Broadbent}, \citenamefont {Howell},\ and\
  \citenamefont {Jordan}}]{DresselPRL2011}%
  \BibitemOpen
  \bibfield  {author} {\bibinfo {author} {\bibfnamefont {J.}~\bibnamefont
  {Dressel}}, \bibinfo {author} {\bibfnamefont {C.~J.}\ \bibnamefont
  {Broadbent}}, \bibinfo {author} {\bibfnamefont {J.~C.}\ \bibnamefont
  {Howell}}, \ and\ \bibinfo {author} {\bibfnamefont {A.~N.}\ \bibnamefont
  {Jordan}},\ }\href {\doibase10.1103/PhysRevLett.106.040402} {\bibfield
  {journal} {\bibinfo  {journal} {Phys. Rev. Lett.}\ }\textbf {\bibinfo
  {volume} {106}},\ \bibinfo {pages} {040402} (\bibinfo {year}
  {2011})}\BibitemShut {NoStop}%
\bibitem [{\citenamefont {Suzuki}\ \emph {et~al.}(2012)\citenamefont {Suzuki},
  \citenamefont {Iinuma},\ and\ \citenamefont {Hofmann}}]{SuzukiNJP2012}%
  \BibitemOpen
  \bibfield  {author} {\bibinfo {author} {\bibfnamefont {Y.}~\bibnamefont
  {Suzuki}}, \bibinfo {author} {\bibfnamefont {M.}~\bibnamefont {Iinuma}}, \
  and\ \bibinfo {author} {\bibfnamefont {H.~F.}\ \bibnamefont {Hofmann}},\
  }\href {http://stacks.iop.org/1367-2630/14/i=10/a=103022} {\bibfield
  {journal} {\bibinfo  {journal} {New Journal of Physics}\ }\textbf {\bibinfo
  {volume} {14}},\ \bibinfo {pages} {103022} (\bibinfo {year}
  {2012})}\BibitemShut {NoStop}%
\bibitem [{\citenamefont {Zhou}\ \emph {et~al.}(2015)\citenamefont {Zhou},
  \citenamefont {Huelga}, \citenamefont {Li},\ and\ \citenamefont
  {Guo}}]{ZhouPRL2015}%
  \BibitemOpen
  \bibfield  {author} {\bibinfo {author} {\bibfnamefont {Z.-Q.}\ \bibnamefont
  {Zhou}}, \bibinfo {author} {\bibfnamefont {S.~F.}\ \bibnamefont {Huelga}},
  \bibinfo {author} {\bibfnamefont {C.-F.}\ \bibnamefont {Li}}, \ and\ \bibinfo
  {author} {\bibfnamefont {G.-C.}\ \bibnamefont {Guo}},\ }\href
  {\doibase10.1103/PhysRevLett.115.113002} {\bibfield  {journal} {\bibinfo
  {journal} {Phys. Rev. Lett.}\ }\textbf {\bibinfo {volume} {115}},\ \bibinfo
  {pages} {113002} (\bibinfo {year} {2015})}\BibitemShut {NoStop}%
\bibitem [{\citenamefont {Waldherr}\ \emph {et~al.}(2011)\citenamefont
  {Waldherr}, \citenamefont {Neumann}, \citenamefont {Huelga}, \citenamefont
  {Jelezko},\ and\ \citenamefont {Wrachtrup}}]{WaldherrPRL2011}%
  \BibitemOpen
  \bibfield  {author} {\bibinfo {author} {\bibfnamefont {G.}~\bibnamefont
  {Waldherr}}, \bibinfo {author} {\bibfnamefont {P.}~\bibnamefont {Neumann}},
  \bibinfo {author} {\bibfnamefont {S.~F.}\ \bibnamefont {Huelga}}, \bibinfo
  {author} {\bibfnamefont {F.}~\bibnamefont {Jelezko}}, \ and\ \bibinfo
  {author} {\bibfnamefont {J.}~\bibnamefont {Wrachtrup}},\ }\href
  {\doibase10.1103/PhysRevLett.107.090401} {\bibfield  {journal} {\bibinfo
  {journal} {Phys. Rev. Lett.}\ }\textbf {\bibinfo {volume} {107}},\ \bibinfo
  {pages} {090401} (\bibinfo {year} {2011})}\BibitemShut {NoStop}%
\bibitem [{\citenamefont {George}\ \emph {et~al.}(2013)\citenamefont {George},
  \citenamefont {Robledo}, \citenamefont {Maroney}, \citenamefont {Blok},
  \citenamefont {Bernien}, \citenamefont {Markham}, \citenamefont {Twitchen},
  \citenamefont {Morton}, \citenamefont {Briggs},\ and\ \citenamefont
  {Hanson}}]{GeorgePNAS2013}%
  \BibitemOpen
  \bibfield  {author} {\bibinfo {author} {\bibfnamefont {R.~E.}\ \bibnamefont
  {George}}, \bibinfo {author} {\bibfnamefont {L.~M.}\ \bibnamefont {Robledo}},
  \bibinfo {author} {\bibfnamefont {O.~J.~E.}\ \bibnamefont {Maroney}},
  \bibinfo {author} {\bibfnamefont {M.~S.}\ \bibnamefont {Blok}}, \bibinfo
  {author} {\bibfnamefont {H.}~\bibnamefont {Bernien}}, \bibinfo {author}
  {\bibfnamefont {M.~L.}\ \bibnamefont {Markham}}, \bibinfo {author}
  {\bibfnamefont {D.~J.}\ \bibnamefont {Twitchen}}, \bibinfo {author}
  {\bibfnamefont {J.~J.~L.}\ \bibnamefont {Morton}}, \bibinfo {author}
  {\bibfnamefont {G.~A.~D.}\ \bibnamefont {Briggs}}, \ and\ \bibinfo {author}
  {\bibfnamefont {R.}~\bibnamefont {Hanson}},\ }\href
  {\doibase10.1073/pnas.1208374110} {\bibfield  {journal} {\bibinfo  {journal}
  {Proceedings of the National Academy of Sciences}\ }\textbf {\bibinfo
  {volume} {110}},\ \bibinfo {pages} {3777} (\bibinfo {year}
  {2013})}\BibitemShut {NoStop}%
\bibitem [{\citenamefont {Athalye}\ \emph {et~al.}(2011)\citenamefont
  {Athalye}, \citenamefont {Roy},\ and\ \citenamefont
  {Mahesh}}]{AthalyePRL2011}%
  \BibitemOpen
  \bibfield  {author} {\bibinfo {author} {\bibfnamefont {V.}~\bibnamefont
  {Athalye}}, \bibinfo {author} {\bibfnamefont {S.~S.}\ \bibnamefont {Roy}}, \
  and\ \bibinfo {author} {\bibfnamefont {T.~S.}\ \bibnamefont {Mahesh}},\
  }\href {\doibase10.1103/PhysRevLett.107.130402} {\bibfield  {journal}
  {\bibinfo  {journal} {Phys. Rev. Lett.}\ }\textbf {\bibinfo {volume} {107}},\
  \bibinfo {pages} {130402} (\bibinfo {year} {2011})}\BibitemShut {NoStop}%
\bibitem [{\citenamefont {Souza}\ \emph {et~al.}(2011)\citenamefont {Souza},
  \citenamefont {Oliveira},\ and\ \citenamefont {Sarthour}}]{SouzaNJP2011}%
  \BibitemOpen
  \bibfield  {author} {\bibinfo {author} {\bibfnamefont {A.~M.}\ \bibnamefont
  {Souza}}, \bibinfo {author} {\bibfnamefont {I.~S.}\ \bibnamefont {Oliveira}},
  \ and\ \bibinfo {author} {\bibfnamefont {R.~S.}\ \bibnamefont {Sarthour}},\
  }\href {http://stacks.iop.org/1367-2630/13/i=5/a=053023} {\bibfield
  {journal} {\bibinfo  {journal} {New Journal of Physics}\ }\textbf {\bibinfo
  {volume} {13}},\ \bibinfo {pages} {053023} (\bibinfo {year}
  {2011})}\BibitemShut {NoStop}%
\bibitem [{\citenamefont {Knee}\ \emph {et~al.}(2012)\citenamefont {Knee},
  \citenamefont {Simmons}, \citenamefont {Gauger}, \citenamefont {Morton},
  \citenamefont {Riemann}, \citenamefont {Abrosimov}, \citenamefont {Becker},
  \citenamefont {Pohl}, \citenamefont {Itoh}, \citenamefont {Thewalt},
  \citenamefont {Briggs},\ and\ \citenamefont {Benjamin}}]{KneSimGau12}%
  \BibitemOpen
  \bibfield  {author} {\bibinfo {author} {\bibfnamefont {G.~C.}\ \bibnamefont
  {Knee}}, \bibinfo {author} {\bibfnamefont {S.}~\bibnamefont {Simmons}},
  \bibinfo {author} {\bibfnamefont {E.~M.}\ \bibnamefont {Gauger}}, \bibinfo
  {author} {\bibfnamefont {J.~J.~L.}\ \bibnamefont {Morton}}, \bibinfo {author}
  {\bibfnamefont {H.}~\bibnamefont {Riemann}}, \bibinfo {author} {\bibfnamefont
  {N.~V.}\ \bibnamefont {Abrosimov}}, \bibinfo {author} {\bibfnamefont
  {P.}~\bibnamefont {Becker}}, \bibinfo {author} {\bibfnamefont {H.-J.}\
  \bibnamefont {Pohl}}, \bibinfo {author} {\bibfnamefont {K.~M.}\ \bibnamefont
  {Itoh}}, \bibinfo {author} {\bibfnamefont {M.~L.~W.}\ \bibnamefont
  {Thewalt}}, \bibinfo {author} {\bibnamefont {Briggs}}, \ and\ \bibinfo
  {author} {\bibfnamefont {S.~C.}\ \bibnamefont {Benjamin}},\ }\href
  {\doibase10.1038/ncomms1614} {\bibfield  {journal} {\bibinfo  {journal}
  {Nature Communications}\ }\textbf {\bibinfo {volume} {3}},\ \bibinfo {pages}
  {606} (\bibinfo {year} {2012})}\BibitemShut {NoStop}%
\bibitem [{\citenamefont {Katiyar}\ \emph {et~al.}(2013)\citenamefont
  {Katiyar}, \citenamefont {Shukla}, \citenamefont {Rao},\ and\ \citenamefont
  {Mahesh}}]{KatiyarPRA2013}%
  \BibitemOpen
  \bibfield  {author} {\bibinfo {author} {\bibfnamefont {H.}~\bibnamefont
  {Katiyar}}, \bibinfo {author} {\bibfnamefont {A.}~\bibnamefont {Shukla}},
  \bibinfo {author} {\bibfnamefont {K.~R.~K.}\ \bibnamefont {Rao}}, \ and\
  \bibinfo {author} {\bibfnamefont {T.~S.}\ \bibnamefont {Mahesh}},\ }\href
  {\doibase10.1103/PhysRevA.87.052102} {\bibfield  {journal} {\bibinfo
  {journal} {Phys. Rev. A}\ }\textbf {\bibinfo {volume} {87}},\ \bibinfo
  {pages} {052102} (\bibinfo {year} {2013})}\BibitemShut {NoStop}%
\bibitem [{\citenamefont {Robens}\ \emph {et~al.}(2015)\citenamefont {Robens},
  \citenamefont {Alt}, \citenamefont {Meschede}, \citenamefont {Emary},\ and\
  \citenamefont {Alberti}}]{RobensPRX2015}%
  \BibitemOpen
  \bibfield  {author} {\bibinfo {author} {\bibfnamefont {C.}~\bibnamefont
  {Robens}}, \bibinfo {author} {\bibfnamefont {W.}~\bibnamefont {Alt}},
  \bibinfo {author} {\bibfnamefont {D.}~\bibnamefont {Meschede}}, \bibinfo
  {author} {\bibfnamefont {C.}~\bibnamefont {Emary}}, \ and\ \bibinfo {author}
  {\bibfnamefont {A.}~\bibnamefont {Alberti}},\ }\href
  {\doibase10.1103/PhysRevX.5.011003} {\bibfield  {journal} {\bibinfo
  {journal} {Phys. Rev. X}\ }\textbf {\bibinfo {volume} {5}},\ \bibinfo {pages}
  {011003} (\bibinfo {year} {2015})}\BibitemShut {NoStop}%
\bibitem [{\citenamefont {Emary}\ \emph {et~al.}(2014)\citenamefont {Emary},
  \citenamefont {Lambert},\ and\ \citenamefont {Nori}}]{EmariNoriRPR2014}%
  \BibitemOpen
  \bibfield  {author} {\bibinfo {author} {\bibfnamefont {C.}~\bibnamefont
  {Emary}}, \bibinfo {author} {\bibfnamefont {N.}~\bibnamefont {Lambert}}, \
  and\ \bibinfo {author} {\bibfnamefont {F.}~\bibnamefont {Nori}},\ }\href
  {http://stacks.iop.org/0034-4885/77/i=1/a=016001} {\bibfield  {journal}
  {\bibinfo  {journal} {Reports on Progress in Physics}\ }\textbf {\bibinfo
  {volume} {77}},\ \bibinfo {pages} {016001} (\bibinfo {year}
  {2014})}\BibitemShut {NoStop}%
\bibitem [{\citenamefont {Palacios-Laloy}\ \emph {et~al.}(2010)\citenamefont
  {Palacios-Laloy}, \citenamefont {Mallet}, \citenamefont {Nguyen},
  \citenamefont {Bertet}, \citenamefont {Vion}, \citenamefont {Esteve},\ and\
  \citenamefont {Korotkov}}]{PalaciosNatPhot2010}%
  \BibitemOpen
  \bibfield  {author} {\bibinfo {author} {\bibfnamefont {A.}~\bibnamefont
  {Palacios-Laloy}}, \bibinfo {author} {\bibfnamefont {F.}~\bibnamefont
  {Mallet}}, \bibinfo {author} {\bibfnamefont {F.}~\bibnamefont {Nguyen}},
  \bibinfo {author} {\bibfnamefont {P.}~\bibnamefont {Bertet}}, \bibinfo
  {author} {\bibfnamefont {D.}~\bibnamefont {Vion}}, \bibinfo {author}
  {\bibfnamefont {D.}~\bibnamefont {Esteve}}, \ and\ \bibinfo {author}
  {\bibfnamefont {A.~N.}\ \bibnamefont {Korotkov}},\ }\href
  {\doibase10.1038/nphys1641} {\bibfield  {journal} {\bibinfo  {journal}
  {Nature Physics}\ }\textbf {\bibinfo {volume} {6}},\ \bibinfo {pages} {442}
  (\bibinfo {year} {2010})}\BibitemShut {NoStop}%
\bibitem [{\citenamefont {Groen}\ \emph {et~al.}(2013)\citenamefont {Groen},
  \citenamefont {Rist\`e}, \citenamefont {Tornberg}, \citenamefont {Cramer},
  \citenamefont {de~Groot}, \citenamefont {Picot}, \citenamefont {Johansson},\
  and\ \citenamefont {DiCarlo}}]{GroenPRL2013}%
  \BibitemOpen
  \bibfield  {author} {\bibinfo {author} {\bibfnamefont {J.~P.}\ \bibnamefont
  {Groen}}, \bibinfo {author} {\bibfnamefont {D.}~\bibnamefont {Rist\`e}},
  \bibinfo {author} {\bibfnamefont {L.}~\bibnamefont {Tornberg}}, \bibinfo
  {author} {\bibfnamefont {J.}~\bibnamefont {Cramer}}, \bibinfo {author}
  {\bibfnamefont {P.~C.}\ \bibnamefont {de~Groot}}, \bibinfo {author}
  {\bibfnamefont {T.}~\bibnamefont {Picot}}, \bibinfo {author} {\bibfnamefont
  {G.}~\bibnamefont {Johansson}}, \ and\ \bibinfo {author} {\bibfnamefont
  {L.}~\bibnamefont {DiCarlo}},\ }\href
  {\doibase10.1103/PhysRevLett.111.090506} {\bibfield  {journal} {\bibinfo
  {journal} {Phys. Rev. Lett.}\ }\textbf {\bibinfo {volume} {111}},\ \bibinfo
  {pages} {090506} (\bibinfo {year} {2013})}\BibitemShut {NoStop}%
\bibitem [{\citenamefont {Ruskov}\ \emph {et~al.}(2006)\citenamefont {Ruskov},
  \citenamefont {Korotkov},\ and\ \citenamefont {Mizel}}]{RuskovPRL2006}%
  \BibitemOpen
  \bibfield  {author} {\bibinfo {author} {\bibfnamefont {R.}~\bibnamefont
  {Ruskov}}, \bibinfo {author} {\bibfnamefont {A.~N.}\ \bibnamefont
  {Korotkov}}, \ and\ \bibinfo {author} {\bibfnamefont {A.}~\bibnamefont
  {Mizel}},\ }\href {\doibase10.1103/PhysRevLett.96.200404} {\bibfield
  {journal} {\bibinfo  {journal} {Phys. Rev. Lett.}\ }\textbf {\bibinfo
  {volume} {96}},\ \bibinfo {pages} {200404} (\bibinfo {year}
  {2006})}\BibitemShut {NoStop}%
\bibitem [{\citenamefont {Williams}\ and\ \citenamefont
  {Jordan}(2008)}]{WilliamsPRL2008}%
  \BibitemOpen
  \bibfield  {author} {\bibinfo {author} {\bibfnamefont {N.~S.}\ \bibnamefont
  {Williams}}\ and\ \bibinfo {author} {\bibfnamefont {A.~N.}\ \bibnamefont
  {Jordan}},\ }\href {\doibase10.1103/PhysRevLett.100.026804} {\bibfield
  {journal} {\bibinfo  {journal} {Phys. Rev. Lett.}\ }\textbf {\bibinfo
  {volume} {100}},\ \bibinfo {pages} {026804} (\bibinfo {year}
  {2008})}\BibitemShut {NoStop}%
\bibitem [{\citenamefont {Wilde}\ and\ \citenamefont
  {Mizel}(2012)}]{WildeFP2012}%
  \BibitemOpen
  \bibfield  {author} {\bibinfo {author} {\bibfnamefont {M.~M.}\ \bibnamefont
  {Wilde}}\ and\ \bibinfo {author} {\bibfnamefont {A.}~\bibnamefont {Mizel}},\
  }\href {\doibase10.1007/s10701-011-9598-4} {\bibfield  {journal} {\bibinfo
  {journal} {Foundations of Physics}\ }\textbf {\bibinfo {volume} {42}},\
  \bibinfo {pages} {256} (\bibinfo {year} {2012})}\BibitemShut {NoStop}%
\bibitem [{\citenamefont {Grangier}\ \emph {et~al.}(1998)\citenamefont
  {Grangier}, \citenamefont {Levenson},\ and\ \citenamefont
  {Poizat}}]{GrangierN1998}%
  \BibitemOpen
  \bibfield  {author} {\bibinfo {author} {\bibfnamefont {P.}~\bibnamefont
  {Grangier}}, \bibinfo {author} {\bibfnamefont {J.~A.}\ \bibnamefont
  {Levenson}}, \ and\ \bibinfo {author} {\bibfnamefont {J.-P.}\ \bibnamefont
  {Poizat}},\ }\href {http://dx.doi.org/10.1038/25059} {\bibfield  {journal}
  {\bibinfo  {journal} {Nature}\ }\textbf {\bibinfo {volume} {396}},\ \bibinfo
  {pages} {537} (\bibinfo {year} {1998})}\BibitemShut {NoStop}%
\bibitem [{\citenamefont {Braginsky}\ \emph {et~al.}(1980)\citenamefont
  {Braginsky}, \citenamefont {Vorontsov},\ and\ \citenamefont
  {Thorne}}]{BraginskyS1980}%
  \BibitemOpen
  \bibfield  {author} {\bibinfo {author} {\bibfnamefont {V.~B.}\ \bibnamefont
  {Braginsky}}, \bibinfo {author} {\bibfnamefont {Y.~I.}\ \bibnamefont
  {Vorontsov}}, \ and\ \bibinfo {author} {\bibfnamefont {K.~S.}\ \bibnamefont
  {Thorne}},\ }\href {\doibase10.1126/science.209.4456.547} {\bibfield
  {journal} {\bibinfo  {journal} {Science}\ }\textbf {\bibinfo {volume}
  {209}},\ \bibinfo {pages} {547} (\bibinfo {year} {1980})}\BibitemShut
  {NoStop}%
\bibitem [{\citenamefont {Sewell}\ \emph {et~al.}(2013)\citenamefont {Sewell},
  \citenamefont {Napolitano}, \citenamefont {Behbood}, \citenamefont
  {Colangelo},\ and\ \citenamefont {Mitchell}}]{SewellNatPhot2013}%
  \BibitemOpen
  \bibfield  {author} {\bibinfo {author} {\bibfnamefont {R.~J.}\ \bibnamefont
  {Sewell}}, \bibinfo {author} {\bibfnamefont {M.}~\bibnamefont {Napolitano}},
  \bibinfo {author} {\bibfnamefont {N.}~\bibnamefont {Behbood}}, \bibinfo
  {author} {\bibfnamefont {G.}~\bibnamefont {Colangelo}}, \ and\ \bibinfo
  {author} {\bibfnamefont {M.~W.}\ \bibnamefont {Mitchell}},\ }\href
  {http://dx.doi.org/10.1038/nphoton.2013.100} {\bibfield  {journal} {\bibinfo
  {journal} {Nat. Photon.}\ }\textbf {\bibinfo {volume} {7}},\ \bibinfo {pages}
  {517} (\bibinfo {year} {2013})}\BibitemShut {NoStop}%
\bibitem [{\citenamefont {Madsen}\ and\ \citenamefont
  {M\o{}lmer}(2004)}]{MadsenPRA2004}%
  \BibitemOpen
  \bibfield  {author} {\bibinfo {author} {\bibfnamefont {L.~B.}\ \bibnamefont
  {Madsen}}\ and\ \bibinfo {author} {\bibfnamefont {K.}~\bibnamefont
  {M\o{}lmer}},\ }\href {\doibase10.1103/PhysRevA.70.052324} {\bibfield
  {journal} {\bibinfo  {journal} {Phys. Rev. A}\ }\textbf {\bibinfo {volume}
  {70}},\ \bibinfo {pages} {052324} (\bibinfo {year} {2004})}\BibitemShut
  {NoStop}%
\bibitem [{\citenamefont {Colangelo}\ \emph {et~al.}(2013)\citenamefont
  {Colangelo}, \citenamefont {Sewell}, \citenamefont {Behbood}, \citenamefont
  {Ciurana}, \citenamefont {Triginer},\ and\ \citenamefont
  {Mitchell}}]{ColangeloNJP2013}%
  \BibitemOpen
  \bibfield  {author} {\bibinfo {author} {\bibfnamefont {G.}~\bibnamefont
  {Colangelo}}, \bibinfo {author} {\bibfnamefont {R.~J.}\ \bibnamefont
  {Sewell}}, \bibinfo {author} {\bibfnamefont {N.}~\bibnamefont {Behbood}},
  \bibinfo {author} {\bibfnamefont {F.~M.}\ \bibnamefont {Ciurana}}, \bibinfo
  {author} {\bibfnamefont {G.}~\bibnamefont {Triginer}}, \ and\ \bibinfo
  {author} {\bibfnamefont {M.~W.}\ \bibnamefont {Mitchell}},\ }\href
  {http://stacks.iop.org/1367-2630/15/i=10/a=103007} {\bibfield  {journal}
  {\bibinfo  {journal} {New J. Phys.}\ }\textbf {\bibinfo {volume} {15}},\
  \bibinfo {pages} {103007} (\bibinfo {year} {2013})}\BibitemShut {NoStop}%
\bibitem [{\citenamefont {Kubasik}\ \emph {et~al.}(2009)\citenamefont
  {Kubasik}, \citenamefont {Koschorreck}, \citenamefont {Napolitano},
  \citenamefont {de~Echaniz}, \citenamefont {Crepaz}, \citenamefont {Eschner},
  \citenamefont {Polzik},\ and\ \citenamefont {Mitchell}}]{KubasikPRA2009}%
  \BibitemOpen
  \bibfield  {author} {\bibinfo {author} {\bibfnamefont {M.}~\bibnamefont
  {Kubasik}}, \bibinfo {author} {\bibfnamefont {M.}~\bibnamefont
  {Koschorreck}}, \bibinfo {author} {\bibfnamefont {M.}~\bibnamefont
  {Napolitano}}, \bibinfo {author} {\bibfnamefont {S.~R.}\ \bibnamefont
  {de~Echaniz}}, \bibinfo {author} {\bibfnamefont {H.}~\bibnamefont {Crepaz}},
  \bibinfo {author} {\bibfnamefont {J.}~\bibnamefont {Eschner}}, \bibinfo
  {author} {\bibfnamefont {E.~S.}\ \bibnamefont {Polzik}}, \ and\ \bibinfo
  {author} {\bibfnamefont {M.~W.}\ \bibnamefont {Mitchell}},\ }\href
  {\doibase10.1103/PhysRevA.79.043815} {\bibfield  {journal} {\bibinfo
  {journal} {Phys. Rev. A}\ }\textbf {\bibinfo {volume} {79}},\ \bibinfo
  {pages} {043815} (\bibinfo {year} {2009})}\BibitemShut {NoStop}%
\bibitem [{\citenamefont {{The {LIGO} Scientific
  Collaboration}}(2009)}]{AbbottNJP2009}%
  \BibitemOpen
  \bibfield  {author} {\bibinfo {author} {\bibnamefont {{The {LIGO} Scientific
  Collaboration}}},\ }\href {http://stacks.iop.org/1367-2630/11/i=7/a=073032}
  {\bibfield  {journal} {\bibinfo  {journal} {New Journal of Physics}\ }\textbf
  {\bibinfo {volume} {11}},\ \bibinfo {pages} {073032} (\bibinfo {year}
  {2009})}\BibitemShut {NoStop}%
\bibitem [{\citenamefont {Kofler}\ and\ \citenamefont
  {Brukner}(2008)}]{KoflerPRL2008}%
  \BibitemOpen
  \bibfield  {author} {\bibinfo {author} {\bibfnamefont {J.}~\bibnamefont
  {Kofler}}\ and\ \bibinfo {author} {\bibfnamefont {{\v{C}}.}~\bibnamefont
  {Brukner}},\ }\href {\doibase10.1103/PhysRevLett.101.090403} {\bibfield
  {journal} {\bibinfo  {journal} {Phys. Rev. Lett.}\ }\textbf {\bibinfo
  {volume} {101}},\ \bibinfo {pages} {090403} (\bibinfo {year}
  {2008})}\BibitemShut {NoStop}%
\bibitem [{\citenamefont {Wang}\ \emph {et~al.}(2013)\citenamefont {Wang},
  \citenamefont {Ghobadi}, \citenamefont {Raeisi},\ and\ \citenamefont
  {Simon}}]{WangPRA2013}%
  \BibitemOpen
  \bibfield  {author} {\bibinfo {author} {\bibfnamefont {T.}~\bibnamefont
  {Wang}}, \bibinfo {author} {\bibfnamefont {R.}~\bibnamefont {Ghobadi}},
  \bibinfo {author} {\bibfnamefont {S.}~\bibnamefont {Raeisi}}, \ and\ \bibinfo
  {author} {\bibfnamefont {C.}~\bibnamefont {Simon}},\ }\href
  {\doibase10.1103/PhysRevA.88.062114} {\bibfield  {journal} {\bibinfo
  {journal} {Phys. Rev. A}\ }\textbf {\bibinfo {volume} {88}},\ \bibinfo
  {pages} {062114} (\bibinfo {year} {2013})}\BibitemShut {NoStop}%
\bibitem [{\citenamefont {Jeong}\ \emph {et~al.}(2014)\citenamefont {Jeong},
  \citenamefont {Lim},\ and\ \citenamefont {Kim}}]{JeongPRL2014}%
  \BibitemOpen
  \bibfield  {author} {\bibinfo {author} {\bibfnamefont {H.}~\bibnamefont
  {Jeong}}, \bibinfo {author} {\bibfnamefont {Y.}~\bibnamefont {Lim}}, \ and\
  \bibinfo {author} {\bibfnamefont {M.~S.}\ \bibnamefont {Kim}},\ }\href
  {\doibase10.1103/PhysRevLett.112.010402} {\bibfield  {journal} {\bibinfo
  {journal} {Phys. Rev. Lett.}\ }\textbf {\bibinfo {volume} {112}},\ \bibinfo
  {pages} {010402} (\bibinfo {year} {2014})}\BibitemShut {NoStop}%
\bibitem [{\citenamefont {Calarco}\ and\ \citenamefont
  {Onofrio}(1995)}]{CalarcoPRA1995}%
  \BibitemOpen
  \bibfield  {author} {\bibinfo {author} {\bibfnamefont {T.}~\bibnamefont
  {Calarco}}\ and\ \bibinfo {author} {\bibfnamefont {R.}~\bibnamefont
  {Onofrio}},\ }\href {\doibase10.1016/0375-9601(94)01020-U} {\bibfield
  {journal} {\bibinfo  {journal} {Physics Letters A}\ }\textbf {\bibinfo
  {volume} {198}},\ \bibinfo {pages} {279 } (\bibinfo {year}
  {1995})}\BibitemShut {NoStop}%
\bibitem [{\citenamefont {Calarco}\ and\ \citenamefont
  {Onofrio}(1997)}]{CalarcoAPB1997}%
  \BibitemOpen
  \bibfield  {author} {\bibinfo {author} {\bibfnamefont {T.}~\bibnamefont
  {Calarco}}\ and\ \bibinfo {author} {\bibfnamefont {R.}~\bibnamefont
  {Onofrio}},\ }\href {\doibase10.1007/s003400050158} {\bibfield  {journal}
  {\bibinfo  {journal} {Applied Physics B}\ }\textbf {\bibinfo {volume} {64}},\
  \bibinfo {pages} {141} (\bibinfo {year} {1997})}\BibitemShut {NoStop}%
\bibitem [{\citenamefont {Ramanathan}\ \emph {et~al.}(2011)\citenamefont
  {Ramanathan}, \citenamefont {Paterek}, \citenamefont {Kay}, \citenamefont
  {Kurzy\ifmmode~\acute{n}\else \'{n}\fi{}ski},\ and\ \citenamefont
  {Kaszlikowski}}]{Ramanathan2007}%
  \BibitemOpen
  \bibfield  {author} {\bibinfo {author} {\bibfnamefont {R.}~\bibnamefont
  {Ramanathan}}, \bibinfo {author} {\bibfnamefont {T.}~\bibnamefont {Paterek}},
  \bibinfo {author} {\bibfnamefont {A.}~\bibnamefont {Kay}}, \bibinfo {author}
  {\bibfnamefont {P.}~\bibnamefont {Kurzy\ifmmode~\acute{n}\else
  \'{n}\fi{}ski}}, \ and\ \bibinfo {author} {\bibfnamefont {D.}~\bibnamefont
  {Kaszlikowski}},\ }\href {\doibase10.1103/PhysRevLett.107.060405} {\bibfield
  {journal} {\bibinfo  {journal} {Phys. Rev. Lett.}\ }\textbf {\bibinfo
  {volume} {107}},\ \bibinfo {pages} {060405} (\bibinfo {year}
  {2011})}\BibitemShut {NoStop}%
\bibitem [{\citenamefont {Budroni}\ \emph {et~al.}(2013)\citenamefont
  {Budroni}, \citenamefont {Moroder}, \citenamefont {Kleinmann},\ and\
  \citenamefont {G\"uhne}}]{BudroniPRL2013}%
  \BibitemOpen
  \bibfield  {author} {\bibinfo {author} {\bibfnamefont {C.}~\bibnamefont
  {Budroni}}, \bibinfo {author} {\bibfnamefont {T.}~\bibnamefont {Moroder}},
  \bibinfo {author} {\bibfnamefont {M.}~\bibnamefont {Kleinmann}}, \ and\
  \bibinfo {author} {\bibfnamefont {O.}~\bibnamefont {G\"uhne}},\ }\href
  {\doibase10.1103/PhysRevLett.111.020403} {\bibfield  {journal} {\bibinfo
  {journal} {Phys. Rev. Lett.}\ }\textbf {\bibinfo {volume} {111}},\ \bibinfo
  {pages} {020403} (\bibinfo {year} {2013})}\BibitemShut {NoStop}%
\bibitem [{\citenamefont {Budroni}\ and\ \citenamefont
  {Emary}(2014)}]{BudroniPRL2014}%
  \BibitemOpen
  \bibfield  {author} {\bibinfo {author} {\bibfnamefont {C.}~\bibnamefont
  {Budroni}}\ and\ \bibinfo {author} {\bibfnamefont {C.}~\bibnamefont
  {Emary}},\ }\href {\doibase10.1103/PhysRevLett.113.050401} {\bibfield
  {journal} {\bibinfo  {journal} {Phys. Rev. Lett.}\ }\textbf {\bibinfo
  {volume} {113}},\ \bibinfo {pages} {050401} (\bibinfo {year}
  {2014})}\BibitemShut {NoStop}%
\bibitem [{\citenamefont {Avis}\ \emph {et~al.}(2010)\citenamefont {Avis},
  \citenamefont {Hayden},\ and\ \citenamefont {Wilde}}]{AvisPRA2010}%
  \BibitemOpen
  \bibfield  {author} {\bibinfo {author} {\bibfnamefont {D.}~\bibnamefont
  {Avis}}, \bibinfo {author} {\bibfnamefont {P.}~\bibnamefont {Hayden}}, \ and\
  \bibinfo {author} {\bibfnamefont {M.~M.}\ \bibnamefont {Wilde}},\ }\href
  {\doibase10.1103/PhysRevA.82.030102} {\bibfield  {journal} {\bibinfo
  {journal} {Phys. Rev. A}\ }\textbf {\bibinfo {volume} {82}},\ \bibinfo
  {pages} {030102} (\bibinfo {year} {2010})}\BibitemShut {NoStop}%
\bibitem [{\citenamefont {{Pitowsky}}(1989)}]{Pitowsky89b}%
  \BibitemOpen
  \bibinfo {editor} {\bibfnamefont {I.}~\bibnamefont {{Pitowsky}}},\ ed.,\
  \href {\doibase10.1007/BFb0021186} {\emph {\bibinfo {title} {Quantum
  Probability - Quantum Logic}}},\ \bibinfo {series} {Lecture Notes in Physics,
  Berlin Springer Verlag}, Vol.\ \bibinfo {volume} {321}\ (\bibinfo
  {publisher} {Springer Verlag Berlin},\ \bibinfo {year} {1989})\BibitemShut
  {NoStop}%
\bibitem [{\citenamefont {Deza}\ and\ \citenamefont {Laurent}(1997)}]{DL97}%
  \BibitemOpen
  \bibfield  {author} {\bibinfo {author} {\bibfnamefont {M.~M.}\ \bibnamefont
  {Deza}}\ and\ \bibinfo {author} {\bibfnamefont {M.}~\bibnamefont {Laurent}},\
  }\href {http://www.ams.org/mathscinet-getitem?mr=1460488} {\emph {\bibinfo
  {title} {{Geometry of cuts and metrics}}}},\ \bibinfo {series} {Algorithms
  and Combinatorics}, Vol.~\bibinfo {volume} {15}\ (\bibinfo  {publisher}
  {Springer-Verlag},\ \bibinfo {address} {Berlin},\ \bibinfo {year}
  {1997})\BibitemShut {NoStop}%
\bibitem [{\citenamefont {Baragiola}\ \emph {et~al.}(2014)\citenamefont
  {Baragiola}, \citenamefont {Norris}, \citenamefont {Monta\~no}, \citenamefont
  {Mickelson}, \citenamefont {Jessen},\ and\ \citenamefont
  {Deutsch}}]{BaragiolaPRA2014}%
  \BibitemOpen
  \bibfield  {author} {\bibinfo {author} {\bibfnamefont {B.~Q.}\ \bibnamefont
  {Baragiola}}, \bibinfo {author} {\bibfnamefont {L.~M.}\ \bibnamefont
  {Norris}}, \bibinfo {author} {\bibfnamefont {E.}~\bibnamefont {Monta\~no}},
  \bibinfo {author} {\bibfnamefont {P.~G.}\ \bibnamefont {Mickelson}}, \bibinfo
  {author} {\bibfnamefont {P.~S.}\ \bibnamefont {Jessen}}, \ and\ \bibinfo
  {author} {\bibfnamefont {I.~H.}\ \bibnamefont {Deutsch}},\ }\href
  {\doibase10.1103/PhysRevA.89.033850} {\bibfield  {journal} {\bibinfo
  {journal} {Phys. Rev. A}\ }\textbf {\bibinfo {volume} {89}},\ \bibinfo
  {pages} {033850} (\bibinfo {year} {2014})}\BibitemShut {NoStop}%
\bibitem [{\citenamefont {Koschorreck}\ and\ \citenamefont
  {Mitchell}(2009)}]{KoschorreckJPB2009}%
  \BibitemOpen
  \bibfield  {author} {\bibinfo {author} {\bibfnamefont {M.}~\bibnamefont
  {Koschorreck}}\ and\ \bibinfo {author} {\bibfnamefont {M.~W.}\ \bibnamefont
  {Mitchell}},\ }\href {http://stacks.iop.org/0953-4075/42/i=19/a=195502}
  {\bibfield  {journal} {\bibinfo  {journal} {Journal of Physics B: Atomic,
  Molecular and Optical Physics}\ }\textbf {\bibinfo {volume} {42}},\ \bibinfo
  {pages} {195502} (\bibinfo {year} {2009})}\BibitemShut {NoStop}%
\bibitem [{\citenamefont {Giedke}\ and\ \citenamefont
  {Cirac}(2002)}]{GiedkePRA2002}%
  \BibitemOpen
  \bibfield  {author} {\bibinfo {author} {\bibfnamefont {G.}~\bibnamefont
  {Giedke}}\ and\ \bibinfo {author} {\bibfnamefont {J.~I.}\ \bibnamefont
  {Cirac}},\ }\href {\doibase10.1103/PhysRevA.66.032316} {\bibfield  {journal}
  {\bibinfo  {journal} {Phys. Rev. A}\ }\textbf {\bibinfo {volume} {66}},\
  \bibinfo {pages} {032316} (\bibinfo {year} {2002})}\BibitemShut {NoStop}%
\bibitem [{\citenamefont {Koschorreck}\ \emph {et~al.}(2010)\citenamefont
  {Koschorreck}, \citenamefont {Napolitano}, \citenamefont {Dubost},\ and\
  \citenamefont {Mitchell}}]{KoschorreckPRL2010b}%
  \BibitemOpen
  \bibfield  {author} {\bibinfo {author} {\bibfnamefont {M.}~\bibnamefont
  {Koschorreck}}, \bibinfo {author} {\bibfnamefont {M.}~\bibnamefont
  {Napolitano}}, \bibinfo {author} {\bibfnamefont {B.}~\bibnamefont {Dubost}},
  \ and\ \bibinfo {author} {\bibfnamefont {M.~W.}\ \bibnamefont {Mitchell}},\
  }\href {\doibase10.1103/PhysRevLett.105.093602} {\bibfield  {journal}
  {\bibinfo  {journal} {Phys. Rev. Lett.}\ }\textbf {\bibinfo {volume} {105}},\
  \bibinfo {pages} {093602} (\bibinfo {year} {2010})}\BibitemShut {NoStop}%
\bibitem [{\citenamefont {Saffman}\ \emph {et~al.}(2009)\citenamefont
  {Saffman}, \citenamefont {Oblak}, \citenamefont {Appel},\ and\ \citenamefont
  {Polzik}}]{SaffmanPRA2009}%
  \BibitemOpen
  \bibfield  {author} {\bibinfo {author} {\bibfnamefont {M.}~\bibnamefont
  {Saffman}}, \bibinfo {author} {\bibfnamefont {D.}~\bibnamefont {Oblak}},
  \bibinfo {author} {\bibfnamefont {J.}~\bibnamefont {Appel}}, \ and\ \bibinfo
  {author} {\bibfnamefont {E.~S.}\ \bibnamefont {Polzik}},\ }\href
  {\doibase10.1103/PhysRevA.79.023831} {\bibfield  {journal} {\bibinfo
  {journal} {Phys. Rev. A}\ }\textbf {\bibinfo {volume} {79}},\ \bibinfo
  {pages} {023831} (\bibinfo {year} {2009})}\BibitemShut {NoStop}%
\bibitem [{\citenamefont {Mitchell}\ \emph {et~al.}(2012)\citenamefont
  {Mitchell}, \citenamefont {Koschorreck}, \citenamefont {Kubasik},
  \citenamefont {Napolitano},\ and\ \citenamefont {Sewell}}]{MitchellNJP2012}%
  \BibitemOpen
  \bibfield  {author} {\bibinfo {author} {\bibfnamefont {M.~W.}\ \bibnamefont
  {Mitchell}}, \bibinfo {author} {\bibfnamefont {M.}~\bibnamefont
  {Koschorreck}}, \bibinfo {author} {\bibfnamefont {M.}~\bibnamefont
  {Kubasik}}, \bibinfo {author} {\bibfnamefont {M.}~\bibnamefont {Napolitano}},
  \ and\ \bibinfo {author} {\bibfnamefont {R.~J.}\ \bibnamefont {Sewell}},\
  }\href {http://stacks.iop.org/1367-2630/14/i=8/a=085021} {\bibfield
  {journal} {\bibinfo  {journal} {New Journal of Physics}\ }\textbf {\bibinfo
  {volume} {14}},\ \bibinfo {pages} {085021} (\bibinfo {year}
  {2012})}\BibitemShut {NoStop}%
\bibitem [{Note1()}]{Note1}%
  \BibitemOpen
  \bibinfo {note} {It is worth recalling that, in this approximation, the
  outcomes have a continuous and unbounded spectrum $y \in \protect \mathbb R$,
  although the total angular momentum is discrete and bounded.}\BibitemShut
  {Stop}%
\bibitem [{Note2()}]{Note2}%
  \BibitemOpen
  \bibinfo {note} {The literature contains different notions of coarse-graining
  \cite {KoflerPRL2008, WangPRA2013, JeongPRL2014,BudroniPRL2014}. We note that
  our procedure of splitting the range of $y$ into two regions is precisely
  that of the original LG proposal \cite {LeggettGargPRL1985}.}\BibitemShut
  {Stop}%
\bibitem [{Note3()}]{Note3}%
  \BibitemOpen
  \bibinfo {note} {The vanishing of $d$ implies low damage, not high
  measurement resolution. The case of finite optical depth will be treated in a
  separate publication.}\BibitemShut {Stop}%
\end{thebibliography}%

\end{document}